# Quantitative and systematic analysis of bias dependence of spin accumulation voltage in a non-degenerate Si spin valve

Soobeom Lee[1], Fabien Rortais[1], Ryo Ohshima[1], Yuichiro Ando[1,*], Shinji Miwa[2,#], Yoshishige Suzuki[2], Hayato Koike[3] and Masashi Shiraishi[1,*]

1. Department of Electronic Science and Engineering, Kyoto Univ., Kyoto 615-8510, Japan
2. Graduate School of Engineering Science, Osaka Univ., Toyonaka 560-8531, Japan
3. Advanced Products Development Center, TDK Corporation, Ichikawa 272-8558, Japan

* Corresponding authors: Yuichiro Ando (ando@kuee.kyoto-u.ac.jp), Masashi Shiraishi (mshiraishi@kuee.kyoto-u.ac.jp)
# Present address: Institute for Solid State Physics, Univ. Tokyo, Kashiwa 277-8581, Japan

**Abstract**

Spin accumulation voltages in a non-degenerate Si spin valve are discussed quantitatively as a function of electric bias current using systematic experiments and model calculations. As an open question in semiconductor spintronics, the origin of the deviation of spin accumulation voltages measured experimentally in a non-degenerate Si spin valve is clarified from that obtained by model calculation using the spin drift diffusion equation including the effect of the spin-dependent interfacial resistance of tunneling barriers. Unlike the case of metallic spin valves, the bias dependence of the resistance-area product for a ferromagnet/MgO/Si interface, resulting in the reappearance of the conductance mismatch, plays a central role to induce the deviation.

# I. Introduction

Silicon (Si) spintronics has been attracting significant attention in the last decade from both fundamental and applied viewpoints [1-17]. Lattice inversion symmetry and the small atomic number of Si allow good spin coherence that has been demonstrated experimentally, and investigations of the spin relaxation mechanism and realization of long-range spin transport has attracted many physicists. These achievements and understandings of basic spin transport and relaxation physics in Si provide a firm basis for the fabrication of novel spin devices such as spin metal-oxide-semiconductors field-effect-transistors (MOSFETs) and spin quantum devices.

A Si spin MOSFET is regarded as one of the most potential candidates as a novel semiconductor-based spin device, and much effort has been paid to the demonstration of Si spin MOSFET operation using non-degenerate Si at room temperature (RT) [11]. After the demonstration, the realization of high on/off spin voltage ratio [12] and of a large spin accumulation voltage of 1.5 mV at RT [14] were achieved in rapid succession. In the study [14], the bias current dependence of spin accumulation voltages in non-degenerate Si with a spin MOSFET structure was studied based on a conventional spin drift diffusion model [18,19] with a trilayer (Fe/Si/Fe) structure and including spin-dependent interfacial resistance due to a tunneling barrier (MgO) between the Fe and Si. The spin voltages as a function of the electric bias current obtained experimentally in degenerate Si (and also Cu) spin valves were well reproduced by the model calculation. However, the dependence of the spin voltages in a non-degenerate Si spin valve still exhibited deviation from the model calculation result under a higher bias current application, and the amplitude of the spin voltages measured experimentally became smaller than that obtained by the calculation. The origin of this deviation remains one of the most important unsolved issues in the physics of spin transport, accumulation and detection in semiconductor spintronics. Bias current or voltage dependence of spin signals in spin devices using a wide variety of materials, such as nonmagnetic metals [20], GaAs [21], Si [22], and graphene [23] has been gaining much attention, where the spin polarization of injected and propagating spins was mainly revealed. However, the previous study on the bias dependence of the spin voltages in Si [22] was implemented only within a scheme of non-local measurement [24], which is a different measurement scheme to that for spin MOSFET operation [11,12]. Source-drain bias current application is indispensable for the operation of Si spin MOSFETs and the amplitude of a spin voltage in a spin MOSFET determines the performance of spin-based logic systems such as reconfigurable logic circuits [1] that consist of Si spin MOSFETs; therefore, an understanding of the bias dependence of the spin voltages in a non-degenerate Si spin device can provide significant information for further progress in studies on Si spin MOSFETs. Thus, to understand the origin of the deviation of the spin voltages as a function of the bias current is important.

Here, we clarify the origin of the remaining inconsistency between the experimentally measured and the theoretically calculated spin accumulation voltages as a function of the bias current. Systematic experiments and modification of the trilayer model allow quantitative analysis of what induced the deviation in the spin accumulation voltages as a function of the bias current. The analysis here in this study provides an answer to the currently significant remaining issue in semiconductor spintronics, a deep insight into how to extract large spin accumulation voltages in semiconductor spin devices, and provision for the suppression of spin accumulation voltages in spin devices, which will enable further progress in semiconductor spintronics.

II. Experimental

Non-degenerate Si-based lateral spin valve (LSV) devices were fabricated on a silicon-on-insulator substrate with a structure of 100 nm thick Si(100)/200 nm thick $SiO_2$/bulk Si(100). Phosphorus (P) was ion-implanted into the 100 nm thick Si(100) channel as an n-type dopant. The dopant concentration in the Si channel estimated using secondary ion mass spectroscopy (SIMS) was ca. $1\times10^{18}$ cm$^{-3}$, which indicates non-degenerate silicon. The conductivity of the Si channel ($\sigma_{Si}$) measured using a conventional four-terminal method was $1.86\times10^3$ $\Omega^{-1}$ m$^{-1}$ at 300 K. Prior to deposition of ferromagnetic metal/tunnel barrier layers, a 20 nm thick highly doped silicon layer was grown by magnetron sputtering to suppress the formation of a depletion layer. Al (3 nm)/Fe (12.4 nm)/Co (0.6 nm)/MgO (0.8 nm) layers were subsequently deposited on the Si channel by molecular beam epitaxy as ferromagnetic (FM) electrodes. After deposition of all the layers, the Si spin channel with two FM contacts was fabricated by electron beam lithography and argon ion milling. The sizes of the two FM contacts were $0.2\times21$ μm$^2$ and $2.0\times21$ μm$^2$. The center-to-center distances between the two Fe contacts were set to be 1.8, 2.0, 2.25 and 2.5 μm, and the LSV with a center-to-center distance of 2.25 μm was used for investigation of the electric bias current dependence of the spin accumulation voltages. Note that the top surface of the Si spin channel was etched to remove 20 nm of the highly doped silicon layer. Finally, two outer nonmagnetic (NM) electrodes were fabricated.

Figures 1(a) and (b) show schematics of the local three-terminal magnetoresistance (L3T-MR) [10] and non-local four-terminal (NL4T) measurements [24] used in this study, respectively. All measurements were performed under direct current (dc) conditions at 300 K. We stress that the L3T-MR measurement is different from the three-terminal Hanle effect measurement [25]; spin polarized electrons in the Si channel propagate by both drift and diffusion in the L3T-MR configuration. The spin accumulation voltage was measured by sweeping an external in-plane magnetic field in both measurement configurations. Figures 1(c) and (d) show typical results of spin-transport-induced magnetoresistance from L3T-MR and NL4T measurements at 300 K under an injection current of 1 mA, where clear spin accumulation voltage signals and resistance hysteresis can be observed.

To estimate the diffusion constant of the Si, the Hall measurements were conducted at 300 K. A Hall bar device with AuSb ohmic contacts on the non-degenerate Si was fabricated using electron beam lithography, argon ion milling and resistance heating deposition. A schematic illustration of the Hall bar is shown in Fig. 2(a). The channel width and thickness were 27 μm and 80 nm, respectively. The Hall voltage ($V_{\text{Hall}}$) was measured by application of an out-of-plane magnetic field and dc electric current. The applied dc current was varied from 0.1 mA to 3.5 mA.

**III. Results and Discussion**

Figure 3 shows the dependence of the measured spin accumulation voltages from L3T-MR measurements ($\Delta V_{\text{L3T}}$, black closed circles) on the electric current ($I_{\text{inj}}$). Spin voltages calculated using the model established in the previous study [14] as a function of $I_{\text{inj}}$ are also shown in the same figure for comparison. The model details and the equation that describes the amplitude of spin voltages including the spin drift diffusion effect are shown in the Supplemental Materials [26]. The deviation between the calculated and measured $\Delta V_{\text{L3T}}$ is evident and it becomes obvious in the higher bias current region (>0.5 mA). The purpose of this work is to clarify the origin of this deviation. Here, note that the model is simply based on a one-dimensional spin drift diffusion equation (i.e., downstream and upstream spin transport are considered in the non-degenerate Si) including interfacial spin-dependent resistance due to the MgO tunneling barrier, whilst the model does not include any modification of spin transport and device parameters as a function of the electric bias current, e.g. the diffusion constant ($D$), the spin diffusion length ($\lambda_s$), the spin lifetime ($\tau_{sf}$), the interfacial resistance, and the spin polarization. In the following paragraphs, we discuss how the parameters are affected or not affected by the bias current.

It is noteworthy that the diffusion constant of Si can be decreased at a high electric field in the region >1000 V/cm, which was studied in noise measurements and by Monte Carlo simulation [27]. Thus, it is significant to determine the diffusion constant of the Si experimentally as a function of the electric bias current within the spin transport scheme in this study. The diffusion constant was estimated from the Hall measurements (see Fig. 2(a)). Supposing that all the dopants were ionized at 300 K, the total dopant concentration of the Si channel of the Hall bar was estimated to be ca. $5 \times 10^{17}$ cm$^{-3}$ over the entire range of the electric current injected into the Hall bar. The electron mobility was estimated to be ca. 200 cm$^2$/Vs when the Hall factor was set to 1.2 for a doping concentration of $5 \times 10^{17}$ cm$^{-3}$ [28]. The Hall factor was assumed to be constant for the bias current in this study. Figure 2(b) shows the dependence of the diffusion constant on the bias current, where the diffusion constant was almost unchanged with the current. The maximum current densities of the spin transport and the Hall measurements were $1.2 \times 10^9$ A/m$^2$ and $1.6 \times 10^9$ A/m$^2$, respectively; therefore, it is concluded that the diffusion constant of the Si channel of the LSV is independent of the electric bias current and no effect such as the Gunn-type behavior of spin polarization [29] takes place in this measurement scheme.

$\lambda_s$, $D$ and $\tau_{sf}$ have the relation $\lambda_s = (D\tau_{sf})^{0.5}$, and $D$ is corroborated to be unchanged within this spin transport measurement scheme. Therefore, if $\lambda_s$ as a function of the bias current is constant, then $\tau_{sf}$ is concluded to also be constant. However, the suppression of $\tau_{sf}$ was determined under a strong electric field application, which was attributed to intervalley phonon scattering [29]. Therefore, a careful and systematic experiment to determine $\lambda_s$ is necessary. To investigate $\lambda_s$ of non-degenerate Si, spin accumulation voltages in the NL4T ($\Delta V_{NL4T}$) were measured at 300 K. $\Delta V_{NL4T}$ for the samples with various center-to-center distances (*d*) from 1.8 μm to 2.5 μm and under dc current ($I_{inj}$) in the range from 0.06 mA to 2.0 mA are plotted in Fig. 4(a). The solid lines in the figure show fitting lines using the exponential function, exp(-$d/\lambda_s$). $\Delta V_{NL4T}$ measured as a function of *d* was well fitted by the function under each bias current condition, which indicates the spin transport in the Si is governed by spin diffusion, as expected. Figure 4(b) shows $\lambda_s$ as a function of $I_{inj}$, where $\lambda_s$ was estimated to be 1.4±0.2 μm at 300 K, and more importantly, was independent of $I_{inj}$. Consequently, it was corroborated that $\tau_{sf}$ of the Si channel in the LSV was unchanged over the entire range of $I_{inj}$ because *D* and $\lambda_s$ were unchanged.

  *D* and $\lambda_s$, resulting in $\tau_{sf}$, were unchanged with the electric bias current; therefore, further experiments were implemented to elucidate the origin of the deviation. Figures 5(a) and (b) show the resistance-area (RA) product of the FM electrodes as a function of the interfacial bias voltage ($V_{int}$), where the injection and detection FM electrodes were 0.2 μm and 2 μm wide, respectively. The $V_{int}$ dependence of the RA product is prominent and the conductance mismatch [30] manifests itself in the high bias region because the spin resistance of the Si (=$\lambda_s/\sigma_{Si}$) was $8 \times 10^{-10}$ Ω·m². Here, we emphasize that the conductance mismatch was circumvented in the device design of the LSV for the zero bias condition. In fact, the RA product at zero bias was two orders of magnitude greater than the spin resistance of the Si, which is sufficient to avoid the conductance mismatch. However, the application of $V_{int}$ induced a decrease of the RA product, which is ascribable to an enhancement of tunneling probability and/or that of thermionic emission of spin carriers via the tunneling barrier. The suppression of the RA product of the FM contacts can induce suppression of spin injection efficiency and spin detection sensitivity, which is the conductance mismatch. Figures 5(c) and (d) show schematics of the measurement setups to investigate the spin injection efficiency and the spin detection sensitivity as a function of $V_{\text{int}}^{\text{NL4T}}$, respectively, where the narrower FM (0.2 μm wide) was selected for spin injection and the wider FM (2 μm wide) was selected for spin detection, as in the case of the L3T-MR measurement (see Fig. 1(a)). Here, $V_{\text{int}}^{\text{NL4T}}$ was calculated from the total applied voltage in the FM/MgO/Si/NM circuit by subtracting the voltage applied to the Si channel because the contact resistance of Si/NM is negligible. In the NL4T measurement, the electric current flows only in the injection circuit (for example, the left side of the LSV, as shown in Fig. 5(c)), and no electric bias field is applied to the detection circuit (for example, the right side of the LSV, as shown in Fig. 5(c)). Hence, the spin injection efficiency or spin detection sensitivity that varies with $V_{\text{int}}^{\text{NL4T}}$ can be individually examined in the measurement setup, as shown in the Figs. 5(c) and (d). Figures 5(e) and (f) show the $V_{\text{int}}^{\text{NL4T}}$ dependence of $\Delta V_{\text{NL4T}}/I_{\text{inj}}(I_{\text{ext}})$ for the spin injection (extraction) scheme, where the amplitude of $\Delta V_{\text{NL4T}}/I_{\text{inj}}(I_{\text{ext}})$ is normalized according to the amplitude at the minimum bias current. Figures 5(e) and (f) show that $\Delta V_{\text{NL4T}}/I_{\text{inj}}(I_{\text{ext}})$ measured with the NL4T configuration exhibits a sizable and monotonic decrease to $V_{\text{int}}^{\text{NL4T}}$, both in the spin injection and the extraction schemes. A similar monotonic decrease of spin signals with respect to the bias voltage has been reported using an Al LSV with an $Al_2O_3$ tunneling barrier, where it was claimed that the decrease was attributed to a decrease of spin polarization of the FM electrode and sign inversion of the spin signal can appear from a simple free-electron model by assuming a parabolic band structure in the FM [20]. However, the bias dependence of the RA product is negligible in metallic LSVs because the conductance mismatch is automatically circumvented (the conductance mismatch is also circumvented to a certain extent in degenerate semiconductor LSVs [26]). The sizable decrease of the RA product measured in this study represents a significant contribution to the spin injection and detection efficiency in non-degenerate Si LSVs because of the reappearance of the conductance mismatch.

Thus, an attempt was made to reproduce the experimental result shown as black closed circles in Fig. 3 by the introduction of $V_{int}$ dependence of the RA product to the calculation model as a possible modification. The result of the calculation (blue closed circles) including the effect of the RA product modulation shows excellent reproducibility of the experimental results (see Fig. 6), so that the deviation of the spin accumulation voltage in the non-degenerate Si LSV is ascribed to the bias dependence of the RA product of the FM/MgO/Si interface. Importantly, most of the deviation of the spin voltages is accountable for the bias dependence of the RA product, which results in the appearance of the conductance mismatch in the high bias region. It is also notable that the effect of the bias dependence of the RA product on the spin device performance has not been considered and discussed in LSVs with a tunneling barrier. For further progress in the development of Si spin MOSFETs, device design that considers the bias dependence of the RA product to circumvent the conductance mismatch with an application of a sufficient source-drain voltage thus becomes indispensable.

## 4. Summary

We have focused on the electric bias current dependence of spin voltages measured in non-degenerate Si LSVs and investigated the origin of the deviation of the measured spin voltages from the results of model calculation based on the spin drift diffusion equation, including the effect of the spin-dependent interfacial resistance of the tunneling barrier under application of a higher bias current. Quantitative and systematic experiments revealed that the diffusion constant, the spin diffusion length and the spin lifetime are unchanged with the electric bias current within the experimental schemes and that the RA product of the FM/MgO/Si interface exhibits a sizable dependence on the interfacial bias voltages. The bias dependence of the RA product was not included in the model calculation conducted in the previous study [14];

however, the modified model calculation that includes this effect reproduced the experimental results very well. In conclusion, the bias dependence of the RA product, i.e., its sizable decrease with the bias voltage, enables the manifestation of the conductance mismatch, which is the origin of the previously observed deviation.

**Figures and figure captions**

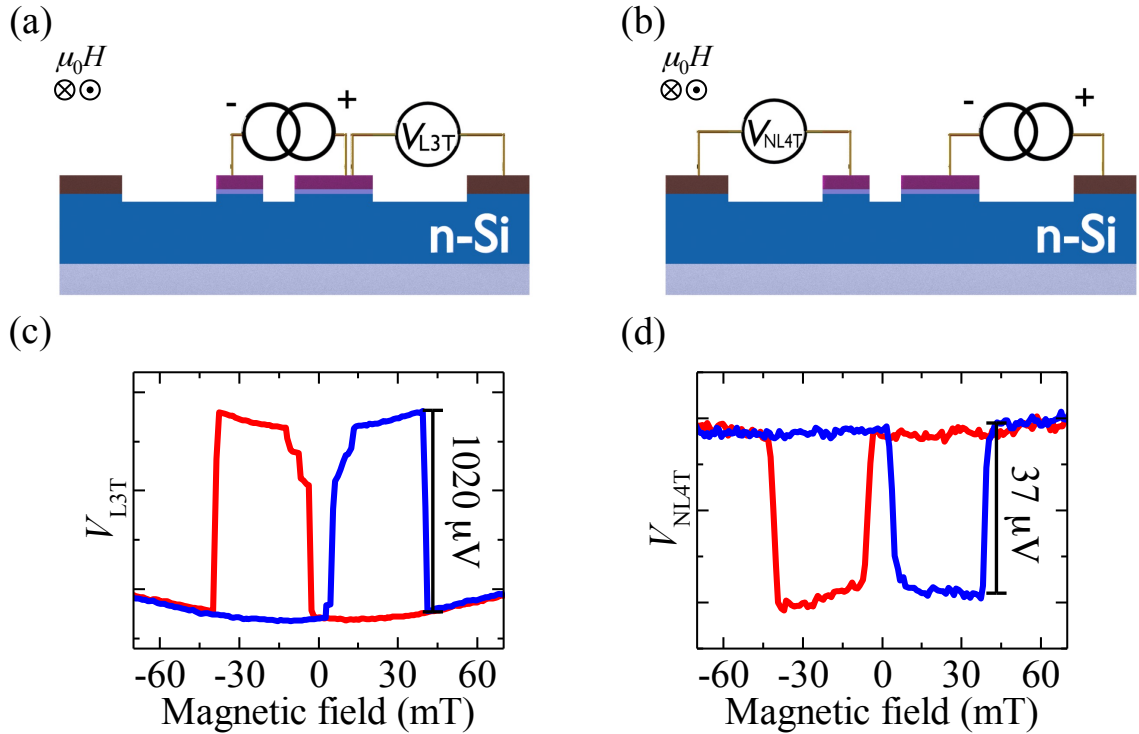

**Figure 1**

(a) Schematic of the measurement for the local three-terminal configuration. Spin polarized electrons were injected at FM1 (left FM electrode; size is 0.2×21 μm$^2$.) and detected at the FM2 (right FM electrode; size is 2.0×21 μm$^2$.). (b) Schematic of the measurement for the non-local four-terminal configuration. Spin polarized electrons are injected at FM2 and detected at FM1. Typical magnetoresistance in (c) the local three-terminal and (d) the non-local four-terminal methods. The center-to-center distance and the applied electric current was 2.25 μm and 1 mA, respectively.

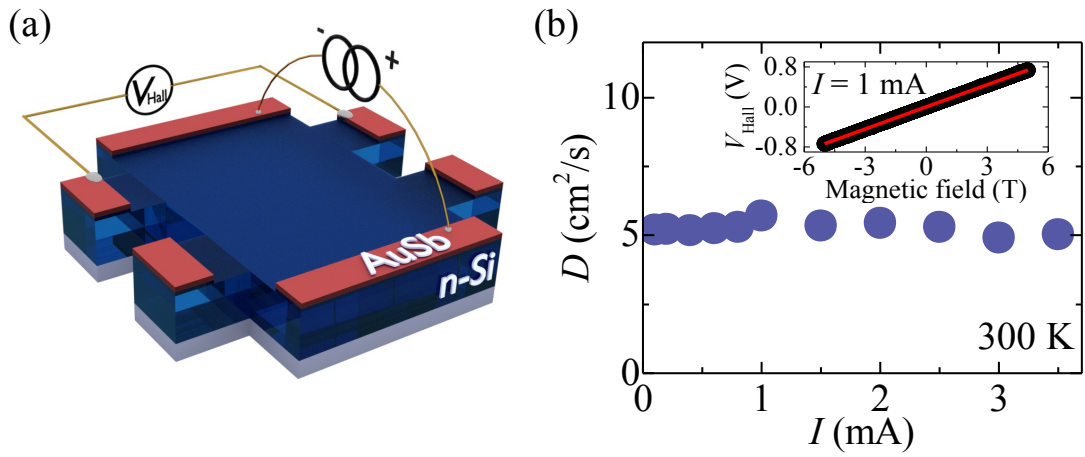

**Figure 2**

(a) Schematic illustration of the Hall bar device of non-degenerate Si with AuSb ohmic contacts.

(b) Current dependence of the diffusion constant in the Si channel. The inset shows an example of the measured Hall voltages at 1 mA.

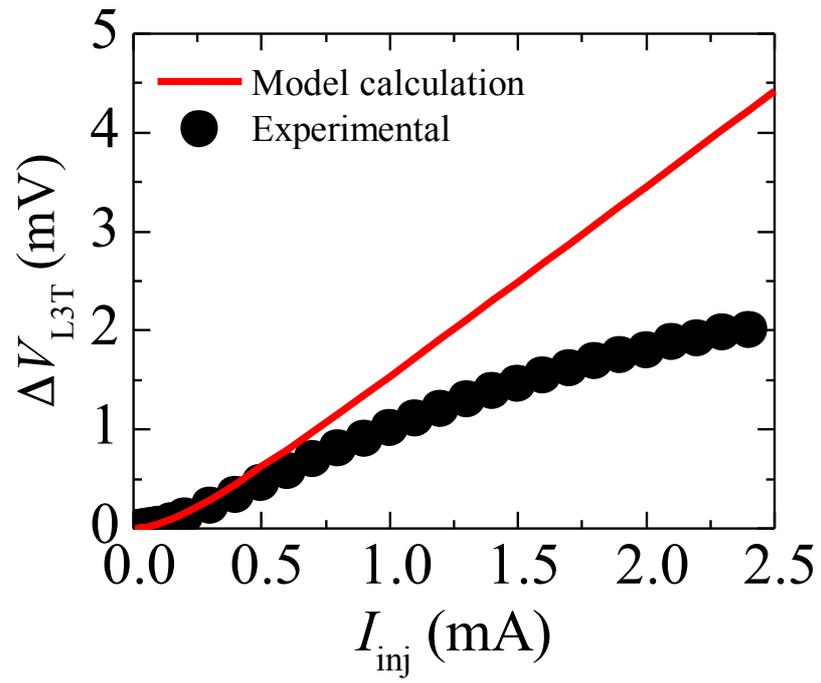

**Figure 3**

Current dependence of theoretically calculated (red line) and experimentally obtained (black dots) $\Delta V_{L3T}$.

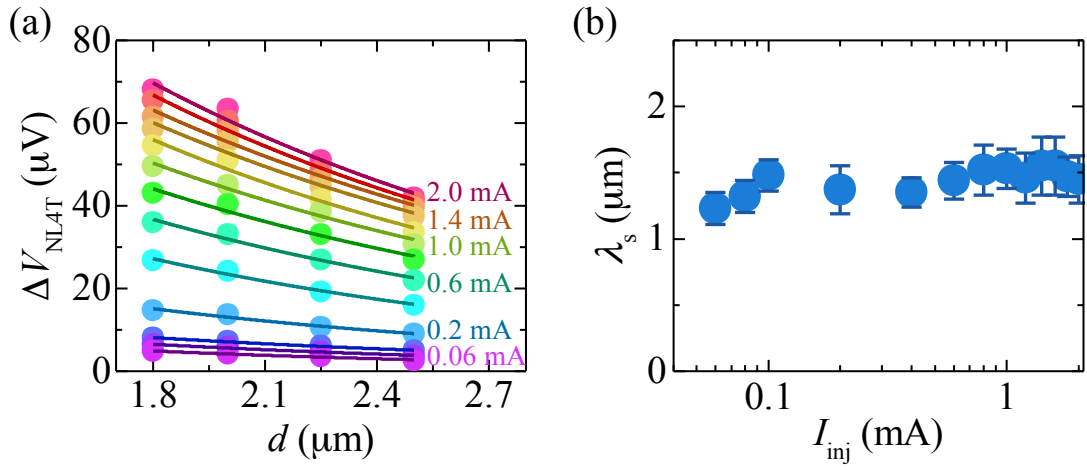

**Figure 4**

(a) Center-to-center distance dependence of $\Delta V_{\text{NL4T}}$ (closed circles) with variation of the magnitude of the electric bias current from 0.06 mA to 2.0 mA. The solid lines are exponential fitting lines. (b) Electric bias current dependence of the spin diffusion length extracted from the results shown in (a).

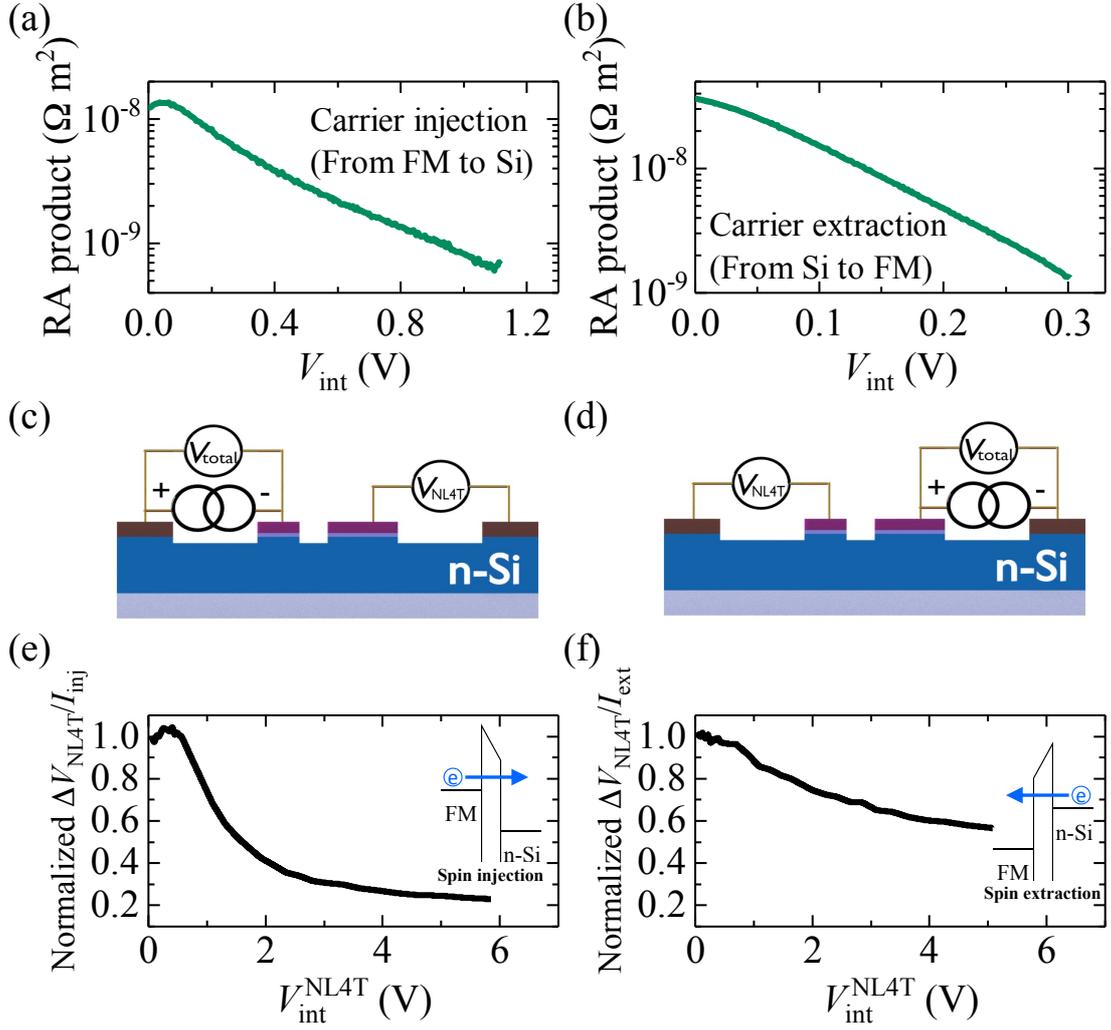

**Figure 5**

Interfacial bias voltage ($V_{int}$) dependence of the RA product of the FM/MgO/Si in the (a) injection and (b) extraction regimes. Non-local four-terminal configurations to examine the variation of (c) spin injection efficiency and (d) spin detection sensitivity for the FM electrodes with $V_{int}^{NL4T}$. Interfacial bias voltage dependence of normalized (e) $\Delta V_{NL4T}/I_{inj}$ and (f) $\Delta V_{NL4T}/I_{ext}$ in injection and extraction regimes, respectively. See the main text for details on the estimation of $V_{int}^{NL4T}$.

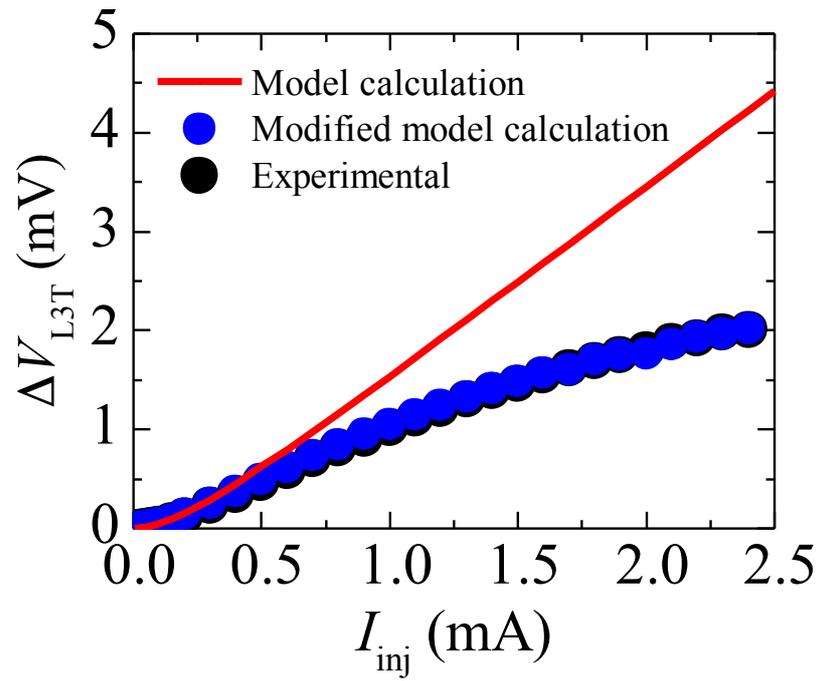

**Figure 6**

Bias current dependence of $\Delta V_{L3T}$. The theoretically calculated (red line) and experimentally obtained (black dots) $\Delta V_{L3T}$ are the same as those shown in Fig. 3, and the blue closed circles show the result of the modified model calculation considering the bias dependence of the RA product.